# The Pie: How Has Human Evolution Distributed Non-Financial Wealth?

*…and what does that mean for financial wealth today?*

Dave Costenaro

*Build Well Consulting, St. Louis, Missouri, United States of America*
Submitted 19 April 2023

## Abstract

Income and wealth allocation are foundational components of how economies and societies operate. These are complex distributions, and it is hard to get a real sense for their characteristics and dynamics using common simplifications such as average or median wealth. One metric that characterizes such distributions better is the Gini Index, which on one extreme is 0, a completely equitable distribution, where everyone has an equal share, and on the other extreme is 1, the most inequitable distribution, where a single individual in the population has all of the resources. Most experts agree that viable economies cannot exist at either extreme, but identifying a preferred value of the Gini Index has historically been a matter of conflicting political philosophies and emotional appeals. This research explores instead whether there might be a theoretical and empirical basis for a preferred Gini Index. More specifically, I explore a simple question: Before commerce and financial systems existed, how were natural assets allocated? Intrinsic human attributes such as height, strength, and beauty were the original measures of value and station in which human social groups traded. Each of these attributes is distributed in a diverse and characteristic way, which I propose has gradually established the acceptable bounds of inequality through evolutionary psychology and human history. I collect data for a wide array of such traits and calculate a Gini Index for them in a novel way assuming their magnitudes to be analogous to levels of financial wealth. The values fall into a surprisingly intuitive pattern ranging from Gini=0.02 to 0.51. Income distributions in many developed countries are within the top half of this range after taxes and transfers are applied (~0.2 to 0.4). Wealth distributions on the other hand are outside of this range in most countries, with the United States at a very inequitable 0.82. Additional research is needed on the interconnections between this range of "natural" Gini Indexes and human contentment; and whether or not any explicit policy goals should be considered to target them.

## Introduction

When studying resource allocation in societies, one sees chart after chart like the one below, where the global share of wealth belongs in an outsized way to a tiny fraction of individuals at the top. So here, in Global wealth distributions, the top 1% of wealthy people on the left bar (represented by the blue segment) have 46% of the total pie of resources on the right bar.

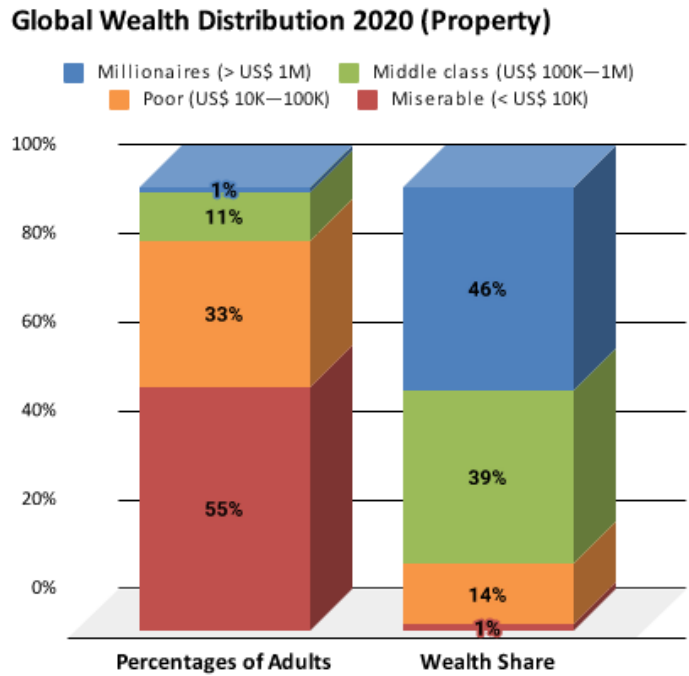

*Figure 1. Global Wealth Distribution 2020 [Image Source: [1]]*

This situation poses an important problem for humanity. Large relative differences in wealth - which are increasingly broadcast around the world 24/7 through social media - strain the fabric of society with perceptions of relative deprivation, a leading psychological cause of depression, mental health issues, and unrest [2]. Additionally, socioeconomic mobility and opportunity are integrally linked to one's neighborhood connections and peer group, so as wealth becomes more concentrated and segregated into fewer neighborhoods, we have fewer thriving neighborhoods acting as springboards when people need economic help [3]. Finally, there is some absolute level of poverty below which people are suffering from basic physical deprivation. Context varies widely, but the World Bank defines extreme poverty generally as living on less than $2.15 per person per day [4].

## On Distributions and the Wealth Allocation Question

Let's look at the spectrum of possible resource distributions by using a couple of pie graphs below. You can think about the pie slices as representing the amount of any resource you want: income, wealth, shares of stock in a company, etc. If you move all the way to the left, there's a completely fair, completely equal distribution where everyone has the same share. As you move to the right, you get shares that have a more highly concentrated distribution. People sometimes associate concepts like socialism or decentralization towards the left of this spectrum, and free markets or centralization toward the right.

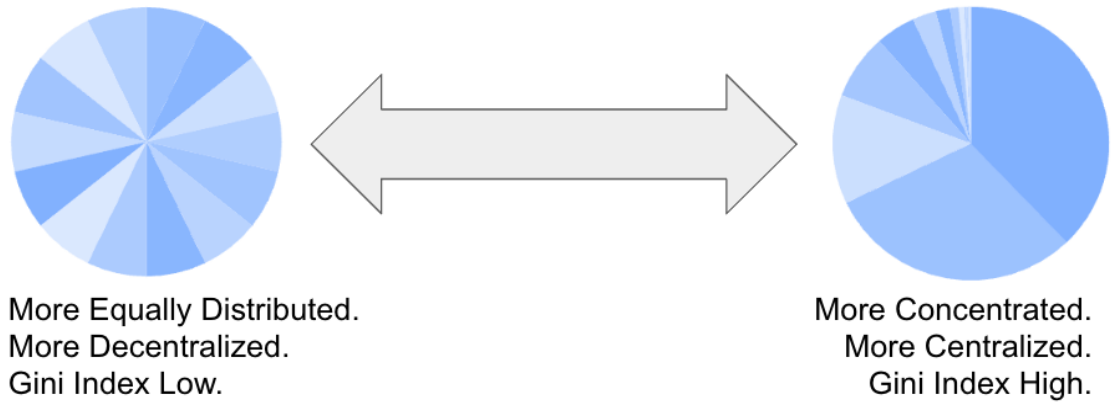

*Figure 2. Spectrum of Resource Distributions*

Are there values on this spectrum that are "correct"? …or at least "acceptable"? This is an enduring question that has been the subject of much philosophical debate, but seemingly with no theoretical or empirical basis. Most reasonable people avoid the extremes and advocate for some fuzzy range in the middle without a rigorous definition.

Proponents of moving leftward on this spectrum advocate that a more equitable economy creates less friction, less conflict, less extreme poverty, and therefore greater happiness and human flourishing. However, proponents of moving rightward on this spectrum argue that the ability to gain a larger share is essential. When there is a ladder of incentives for people to work harder, be more productive, and innovate, then people have the ingredients for personal opportunity and fulfillment. Moreover, larger fortunes have a role in allowing sufficient investments to be made in big, ambitious projects that may not otherwise be possible. These innovations and large scale projects can grow the absolute size of the pie for everyone. In other words, a rising tide lifts all boats, and there is therefore greater happiness and human flourishing.

## The Gini Index

As we have seen, the nature of resource distributions is both important and complex. With this complexity, it is hard to get a working sense of their characteristics and dynamics across an entire population using common simplifications such as average or median. One metric that characterizes resource distributions better is the Gini Index (or Gini Coefficient). Gini Index is calculated by drawing a Lorenz Curve as shown in Figure 3 below. That is, by stacking all of the individuals in a population along the x-axis in ascending-rank order according to the amount of resource they own, which is correspondingly shown on the y-axis.

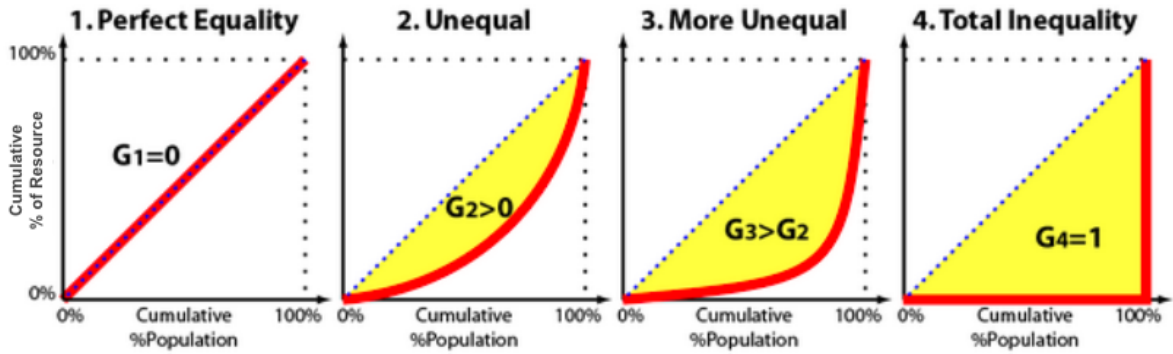

*Figure 3. The Lorenz Curve, a Graphical Representation of Gini Index [Image Source: 5]*

If each new person that is added to the population along the horizontal axis has the same share as the person before them, adding their share on the y-axis will result in a line with a slope of 1, or 45 degrees. This red line represents the perfectly equal distribution shown in step 1 of the figure. If, however, the amount held by each incremental person becomes larger and larger, this will trace a line whose slope of resources is more shallow for those with less share at the left/bottom of the distribution, and it will incrementally slope higher with increasing resources as it advances through the population to the right/top of the distribution, as shown in steps 2 and 3. The extreme then, depicted as step 4 of the figure, is when all of the people stack along the x-axis at a height of zero because they have no resource share, and the last individual has 100% of the resource.

The Gini Index is calculated as the ratio of yellow area divided by the total area under the 45 degree "Perfect Equality" curve.

## How might Evolutionary Psychology inform Wealth Allocation?

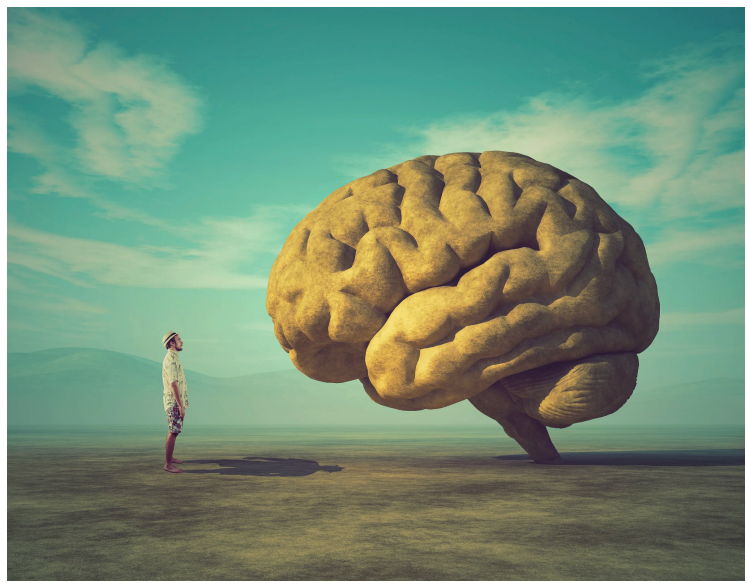

*Figure 4. Image Credit: Areo Magazine*

This research explores a simple question: Before commerce and financial systems existed, how were natural assets allocated? How have human history and evolutionary psychology created an equilibrium in which we regularly navigate our everyday lives, despite a stunning array of human diversity, or in other words: inequality?

I hypothesize that, upon closer inspection, we will find that limits have emerged that naturally bound the level of diversity or inequality in the human condition. Consider a universe where the tallest humans grow to be 10 feet tall, or even 100 feet tall. It is unlikely that the shortest humans would be able to compete, and that the whole distribution would skew taller as those shorter genes died away. On the other hand, there are also limits preventing people from getting that tall in the first place, such as nutrition or the physics of bone density with respect to gravity. Our bodies have also been evolutionarily sculpted by other factors like optimal thermodynamics for metabolism, optimal reproductive and birthing mechanics, optimal performance in care-taking, manual labor, or combat. A whole host of complex interactions have hemmed in this multivariate optimization problem to the N-dimensional probability field that represents an acceptable solution space for any new human.

Let's pursue this concept one layer up from the biological human, now to a social system of many humans. In this context, the number of one's relationships among family and tribe also functions as a kind of asset or wealth, similar to superior height and strength as mentioned above. If one can tap a larger network of relationships for help, this is of higher worth than a smaller network.

## Methodology

For the purposes of this study, I collected and assembled population distribution data for as many such value-imbued human attributes that I could find. I focused on data that was easily quantified, available from reputable sources in sufficient granularity, and served as a reasonable proxy for some societal asset, value, worth, or station. I then calculated the Gini Index for each quantity in a novel way assuming their magnitudes are analogous to a financial asset such as Dollars or Euros. The units of measurement do not matter since the calculations are normalized to use the cumulative *percent* of population and cumulative *percent* of the resource. This means that a Gini Index calculated with financial assets denominated in dollars or euros will be identical, and a Gini Index calculated with height assets denominated in meters or inches will be identical.

The "natural" human assets used in this study are those below:
- 'Height (cm, global adult males)',  [6], [7], [8]
- 'Height (cm, global adult females)',  [6], [7], [8]
- 'IQ'  [9], [10]
- 'Weight (lbs, US adult males)',  [11], [12]
- 'Weight (lbs, US adult females)',  [11], [12]

- 'Average Beauty Score (1 to 5) (OK Cupid Survey)', [13]
- 'Number of Digits Memorized (Short term memory testing)', [14], [15], [16]
- 'Life Expectancy (years, US 2020)', [17], [18]
- 'Life Expectancy (years, global 2020)', [17], [18]
- 'Life Expectancy (years, global 1950)', [17], [18]
- 'Running Speed (mph, global adults)', [19], [20]
- 'Max Benchpress (lbs, global adults)', [21], [22], [23]
- 'Relationships (# Facebook Friends 2015)', [24]

These distributions and resulting Gini Indexes are sensitive to the granularity of the source data, and also vary somewhat from year to year or between data sources where multiple sources were available. There are also questions around how you draw the boundary around the problem. The ideal characterization is: any population or community that is experiencing the given distribution and reaching their mutual equilibrium within it… But does that differ significantly from nation to nation, people group to people group? To the extent possible, I attempted to pick stationary and consistent distributions with data that was representative of all people; or where applicable, sufficiently defined and bounded to the community being measured.

Most of the available data with sufficient granularity to create a population distribution are from modern sources, so the linkage to an evolutionary process also assumes some amount of stability in the distributions over longer periods of time. Some drift is able to be measured, but it does not appear to be sufficient to change the overall trends and conclusions of the study. For example, there is reliable 1950 and 2020 data for the metric "number of years expected to live" [17], [18]. This encapsulates a large change during the last century due to the advent of modern medicine and a corresponding rapid decline in infant mortality. This effectively means that the 1950 distribution has a much larger cohort of the population who die at birth, which appears on the Lorenz Curve as a much larger cohort with no resources.

Despite any caveats to methodology, the results are striking, robust, and easily discernible for the purposes of this study, even assuming one significant digit of precision.

An interesting data point in support of stability of these distributions is around the "number of relationships" metric, which only became measurable for whole population distributions at scale with the advent of modern social networks and data analysis. An analysis of Facebook data from 2015 showed that the average number of Facebook Friends was remarkably close to the famous "Dunbar's Number" of 150, which Dunbar had postulated as an equilibrium network size for humans based on ethnographic research and brain-to-body-weight trends observed in the animal kingdom. He concluded that the technological accelerant of social networks did not increase humans' actual network size in a statistically significant way, postulating that real relationships had other limits such as occasional direct interaction [25].

## Results

Plotted in Figure 5, you can see the Gini Index for the World and select nations. Income is shown in yellow, ranging from Gini=0.47 to 0.68. Wealth is shown in orange, ranging from Gini=0.52 to 0.85 [26], [27], [28].

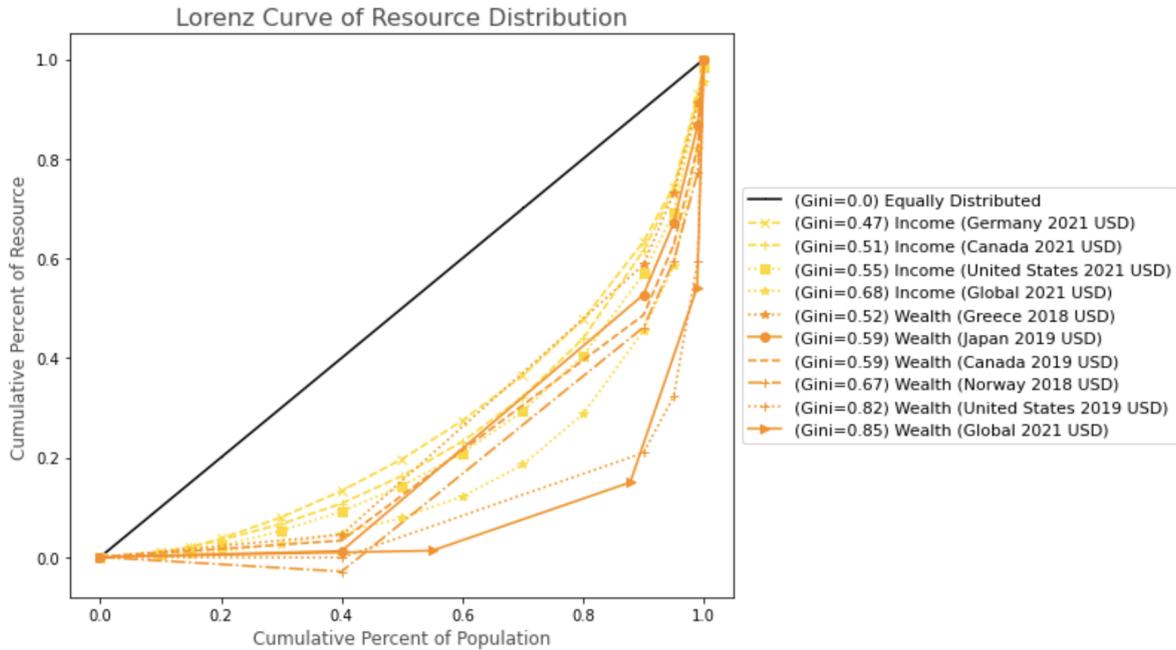

*Figure 5. Lorenz Curve of Income and Wealth Distribution*

Now in Figure 6, we add the Gini Indexes calculated for the "natural" human assets in blue onto the same plot of income and wealth inequality. It is clearly seen that the natural assets occupy a distinct and more equitable space on the graph, closer to the 45 degree curve of equality, but with a fair amount of variation from Gini=0.02 to 0.51.

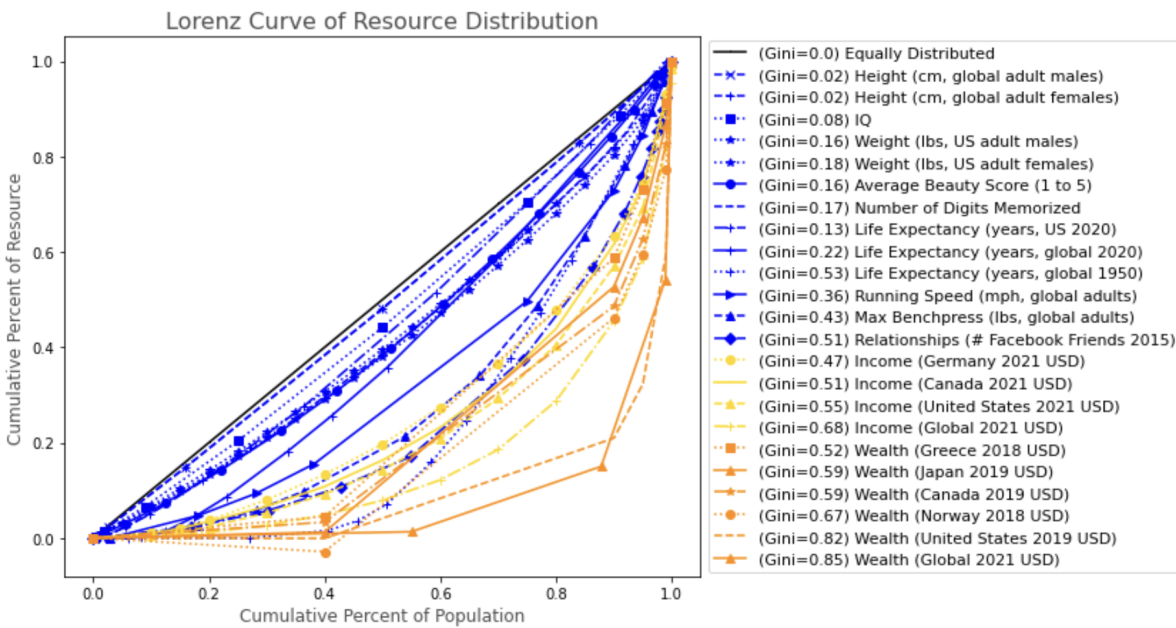

*Figure 6. Lorenz Curve of "Natural" Assets, Income, and Wealth Distribution*

## On Changeable vs Permanent Traits

One very interesting trend is that increases in Gini Index among the natural assets tends to correlate with an increase in changeability, i.e. the degree to which a person can train, hone, or develop a given metric themselves. We talked earlier about the importance of being able to strive and advance up the ladder of a given resource to achieve a larger share. One cannot do this to change their height or IQ since these traits are some of the most permanent ones. They also have the lowest Gini Index. Things like life expectancy, weight, or beauty also have a strong innate/genetic component, but can be meaningfully varied by changes that people make in their lifestyle. Short term memory of digits can also be improved to some degree with focus and training. Toward the top of the Natural Gini range are running speed, maximum benchpress, and number of Facebook Friends; all of which are highly malleable traits where the extreme top percentiles can attain substantially higher levels with dedicated focus. See Figure 7.

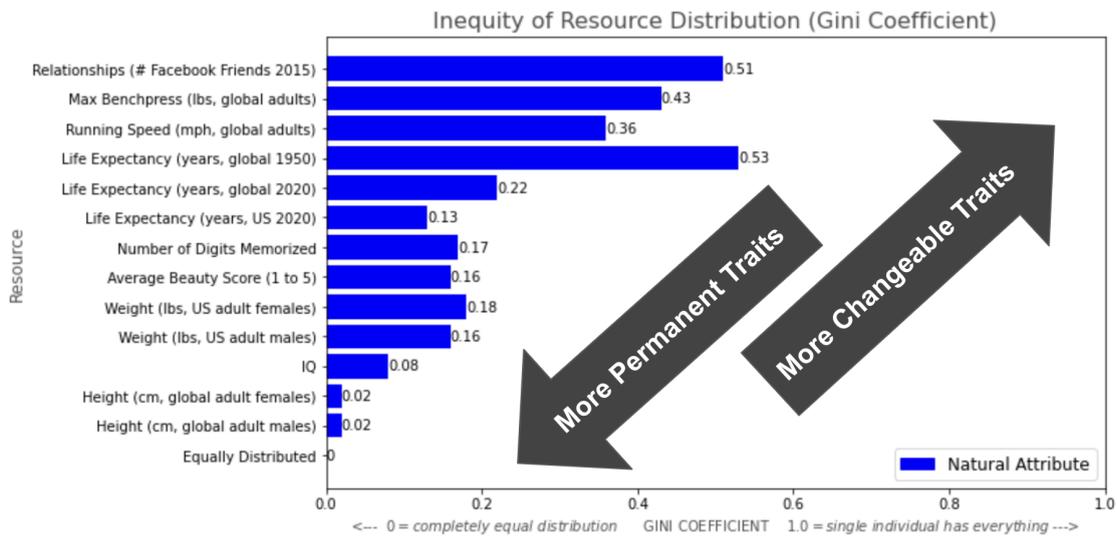

*Figure 7. Natural Asset Correlation between Gini & Changeability*

It is still the case, however, that the malleability and diversity of these natural human assets seem to be bounded to populational distributions with a maximum Gini Index around 0.5.

## On Taxes and Transfers

Figure 8 below adds select income distributions after taxes and transfers have been made. Until now, all income statistics shown have been gross income, and are as such unless otherwise noted. Social welfare and income inequality are clearly already high priorities for governments around the world, and there are systems of taxation and transfers such as welfare programs, tax credits, and the like which attempt to reallocate income to help. For example, the Gini Index for income in the US changes from 0.55 to 0.38 when taxes and transfers are applied, putting income inequality squarely within the hypothesized "natural" range.

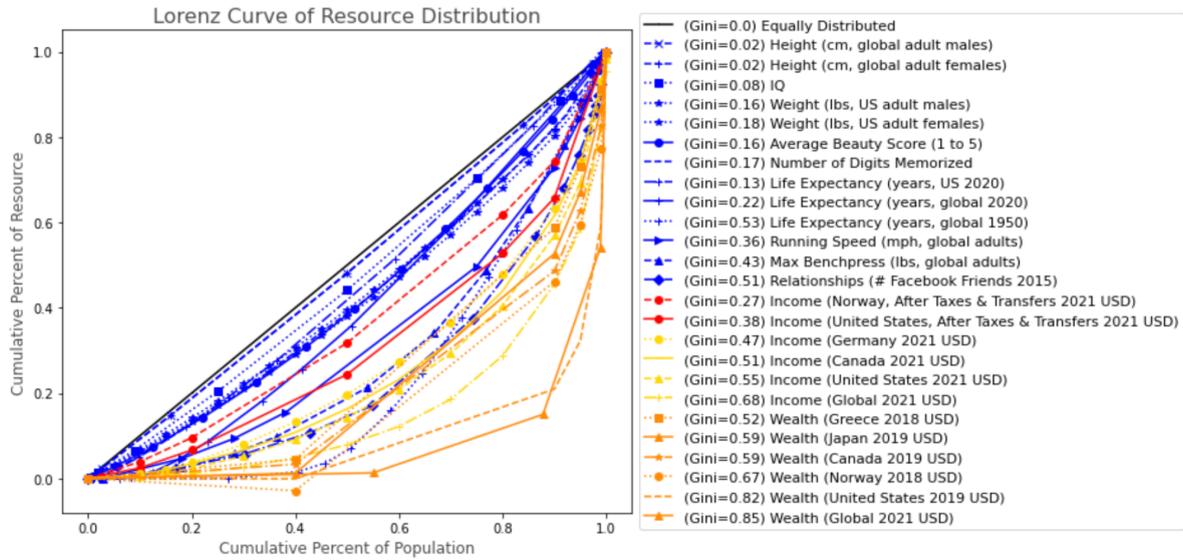

*Figure 8. Lorenz Curve adding Income After Taxes & Transfers*

In fact, for most developed countries, the Gini Index for gross income is around 0.4 to 0.6, but when governments redistribute it through progressive taxation, social programs, and the like, the adjusted Gini Index is more in the 0.2 to 0.4 range [29], again putting it into the range of acceptable "natural" equilibrium hypothesized in this paper.

If income inequality does not "feel" like it is in equilibrium, perhaps there is a qualitative difference between gross income and income after taxes and transfers. Said another way, a larger paycheck is most certainly preferable to food stamps and a tax refund at the end of the year.

The bigger issue, however, may be that wealth inequality has more influence on the human condition than income inequality. Wealth inequality is conspicuously outside the hypothesized equilibrium range, and by its very nature is a more long lasting and impactful attribute than a single year's salary. Distilling the previous findings to show only the Gini Index in Figure 9, we see that wealth allocation in the US has a quite inequitable Gini Index of 0.82. Globally this is marginally worse at 0.85. Greece is the only nation in this study whose wealth Gini Index (0.52) approaches the equilibrium range.

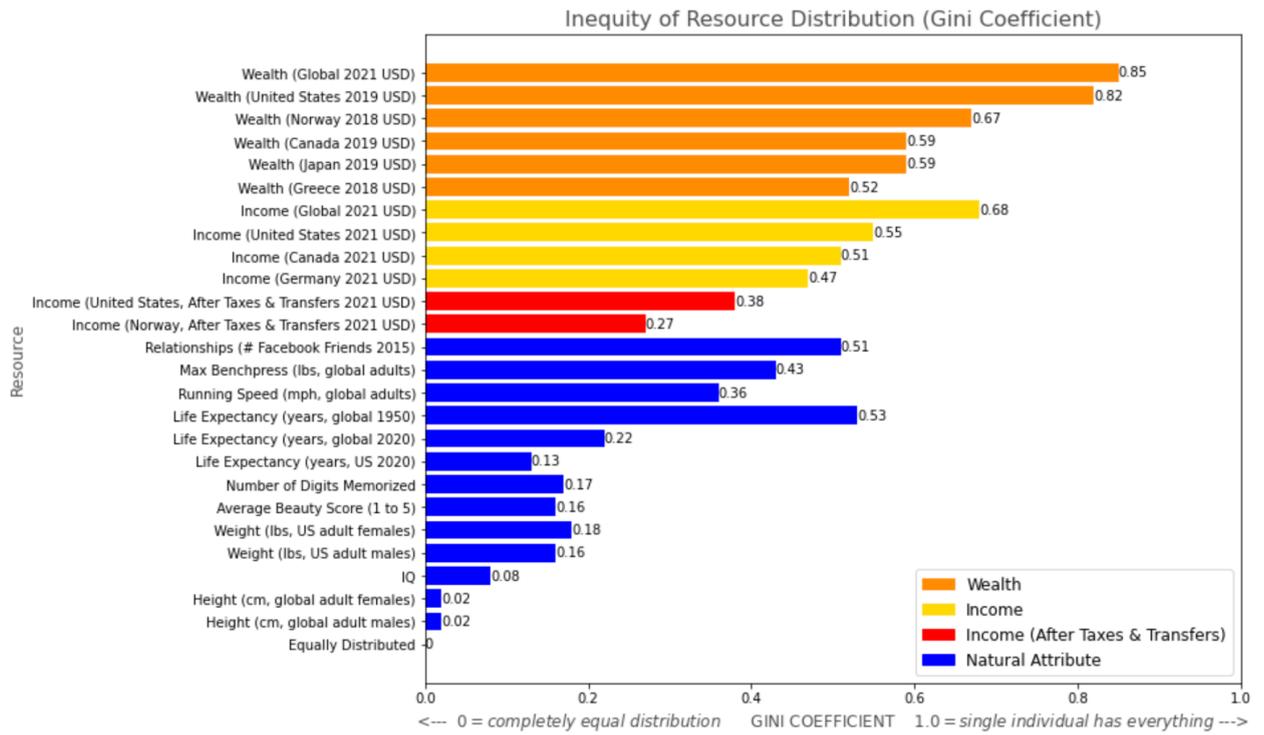

*Figure 9. Bar Graph of Gini Indexes*

## On Real Peoples' Pocketbooks

Let's see how all this translates to real people. We can abstract this data to 100 representative people in a table, showing what the distributions mean for a holistic, financial individual at the various levels in the distribution. I show this is in terms of both their wallet (income) and their assets (wealth).[1] In Figures 10 and 11 below, you can look at the average income and wealth for a person in the given segments for each of three scenarios: the actual current state, a universe in which all the resources are equally distributed, and finally, a universe in which the allocations have been adjusted to result in a Gini Index of 0.5 (on the high end, but still within the hypothesized range of evolutionary equilibrium).[2]

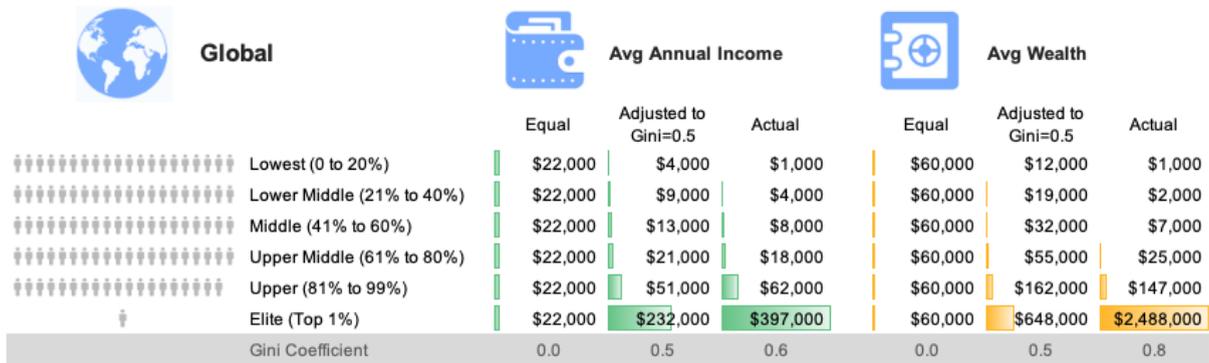

Figure 10. Per Capita Income and Wealth Scenarios, Global

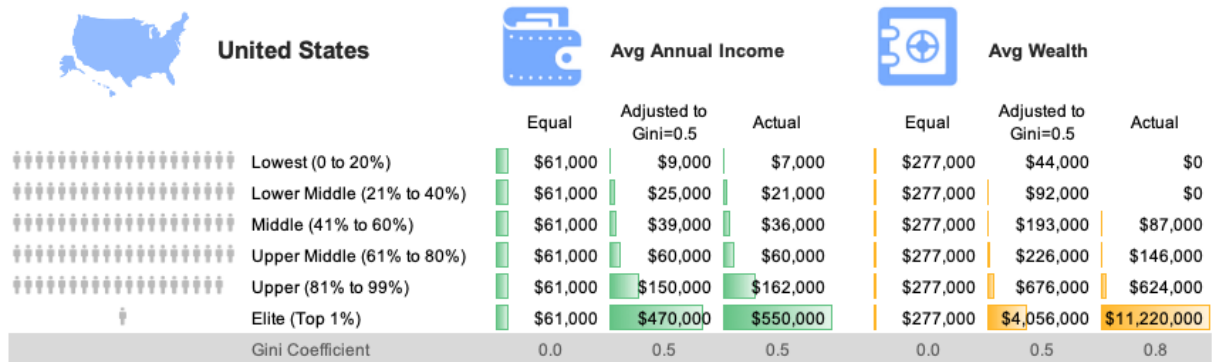

Figure 11. Per Capita Income and Wealth Scenarios, United States

---

[1] Note that these data are from two different distributions, so may not necessarily be the same people, but for simplicity, I assume someone in a given income bracket is also in the same corresponding wealth bracket.

[2] Note that there are many different distribution shapes that can result in the same Gini Index, so I am just presenting one. The methodology used was to increase the resource level of the lowest population segment, and then modify all step changes between subsequent segments by multiplying the ratio from the original distribution by the same factor until the distribution was calibrated to sum up to an equal amount of total resources. Then repeat these steps until the desired Gini Index was achieved.

One can see that a Gini Index adjusted downward in this way could result in income distributions that are somewhat more equal than their current state, both Globally and in the United States. It would, however, result in a substantial shift in wealth and a significantly more equitable wealth distribution, which would allow the poorer segments to have a nest egg and some higher level of stability and savings.

## Conclusions

What are the implications of all this? The observed range of Gini Index when applied to the inequality found in the examined human assets shows a range of 0.02 to 0.51, with more permanent traits at the bottom and more changeable traits at the top of that range.

Is it the case that this might be an empirical recommendation for equilibrium, embedded into the human condition by our evolutionary past? Or does the world of finance extend beyond this pattern, where the gains to innovation or training or personal pursuit, and the accompanying inequality of outcomes, are not bound in the same way?

If we believe that these ranges do, in fact, suggest natural limits to the levels of inequality optimal for human acceptance and flourishing, it could mean a profound reordering of our public policy. Many of the income distributions on an annual basis are already within the observed range after taxes and transfers, but gross income tends to be less so, and wealth distributions are decisively not.

What are possible ways to address this if we wanted to? Wealth taxes are notoriously difficult to define and implement. Perhaps modern technology is now at a point where valuation, transacting, tracking, and implementing such a tax is more feasible. Perhaps reform in estate tax is another route. Or possibly implementing more progressive taxes on real estate and property of higher value. Specific policy solutions, however, are beyond the scope of this study. Substantial additional research is needed on the interconnections between "natural" Gini Indexes and human contentment, and whether or not any explicit policy goals should be considered to target them, but it is my hope that this research provides some helpful framework and insight for these efforts.

*All data and code is open source and available here on GitHub: [https://github.com/dmc314/the-pie](https://github.com/dmc314/the-pie)*